\documentclass[aps, pre, twocolumn, showpacs, amsmath, amssymb, superscriptaddress, reprint, longbibliography]{revtex4-1}
\usepackage[utf8]{inputenc}
\usepackage[T1]{fontenc}
\usepackage{xcolor}
\usepackage{graphicx}
\usepackage{wasysym}

\definecolor{darkblue}{rgb}{0,0,0.6}
\definecolor{darkred}{rgb}{0.6,0,0}
\usepackage[colorlinks=true,urlcolor=darkblue, citecolor=darkblue, linkcolor=darkred]{hyperref}

\newcommand{\ind}[1]{_{\mathrm{#1}}}

\newcommand{\dd}{\mathrm{d}}
\newcommand{\ed}{\mathrm{e}}

\newcommand{\kT}{k_\mathrm{B}T}

\newcommand{\tA}{\tilde A}

\DeclareMathOperator{\erfc}{erfc}

\begin{document}

\title{Unbinding transition of probes in single-file systems}

\author{Olivier Bénichou}
\affiliation{Laboratoire de Physique Théorique de la Matière Condensée, CNRS/UPMC, 4 Place Jussieu, F-75005 Paris, France}

\author{Vincent Démery}
\affiliation{Gulliver, CNRS, ESPCI Paris, PSL Research University, 10 rue Vauquelin, Paris, France}
\affiliation{Univ Lyon, ENS de Lyon, Univ Claude Bernard Lyon 1, CNRS, Laboratoire de Physique, F-69342 Lyon, France}

\author{Alexis Poncet}
\affiliation{Laboratoire de Physique Théorique de la Matière Condensée, CNRS/UPMC, 4 Place Jussieu, F-75005 Paris, France}
\affiliation{Département de Physique, ENS, PSL Research University, 24 Rue Lhomond, 75005 Paris, France}

\begin{abstract}
Single-file transport, arising in quasi one-dimensional geometries where particles cannot pass each other,  is characterized by the anomalous dynamics of a probe, notably its response to an external force.
In these systems, the motion of several probes submitted to different external forces, although relevant to mixtures of charged and neutral or active and passive objects, remains unexplored.
Here, we determine how several probes respond to external forces.
We rely on a hydrodynamic description of the symmetric exclusion process to obtain exact analytical results at long times.
We show that the probes can either move as a whole, or separate into two groups moving away from each other.
In between the two regimes, they separate with a different dynamical exponent, as $t^{1/4}$.
This unbinding transition also occurs in several continuous single-file systems and is expected to be observable.
\end{abstract}

\maketitle

Single-file transport arises in systems as varied as ionic channels~\cite{Finkelstein1981}, nanotubes~\cite{Hummer2001, Berezhkovskii2002, Kalra2003, Tokarz2005, Tunuguntla2017}, and zeolites~\cite{Kukla1996}.
The hallmark of these systems does not lie in the collective dynamics, which is simply diffusive, but in the motion of individual probes~\cite{Levitt1973, Wei2000}.
In absence of external force, a single probe diffuses anomalously due to the interactions with its neighbors, its mean squared displacement scaling as $\langle X(t)^2 \rangle\sim\sqrt{t}$~\cite{Kukla1996, Wei2000, Meersmann2000, Kollmann2003,Lutz2004, Lin2005}. 
In response to a constant external force, its displacement evolves as $X(t)\sim\sqrt{t}$, in agreement with the fluctuation-dissipation theorem~\cite{Burlatsky1996, Landim1998}.
The probability density function of the probe is Gaussian at long times. Finite time corrections have recently been determined~\cite{Krapivsky2014}, and generalized to systems with an initial density gradient~\cite{Imamura2017} or to a driven probe~\cite{Cividini2016b, Kundu2016}.
Remarkably, a driven probe drags with it the surrounding particles, which can be seen as ``bound'' to the probe~\cite{Cividini2016b}.

In contrast, the effects involving several driven probes remain unexplored, despite their relevance to the situations described above.
An especially important example concerns the ionic transport through subnanometer carbon nanotubes~\cite{Tunuguntla2017}.
Indeed, in these nanotubes, water molecules are confined to a single-file chain~\cite{Hummer2001, Tunuguntla2017}, and ions act as driven particles if a potential difference is applied between the reservoirs.

\begin{figure*}
\begin{center}
\includegraphics[scale=.9]{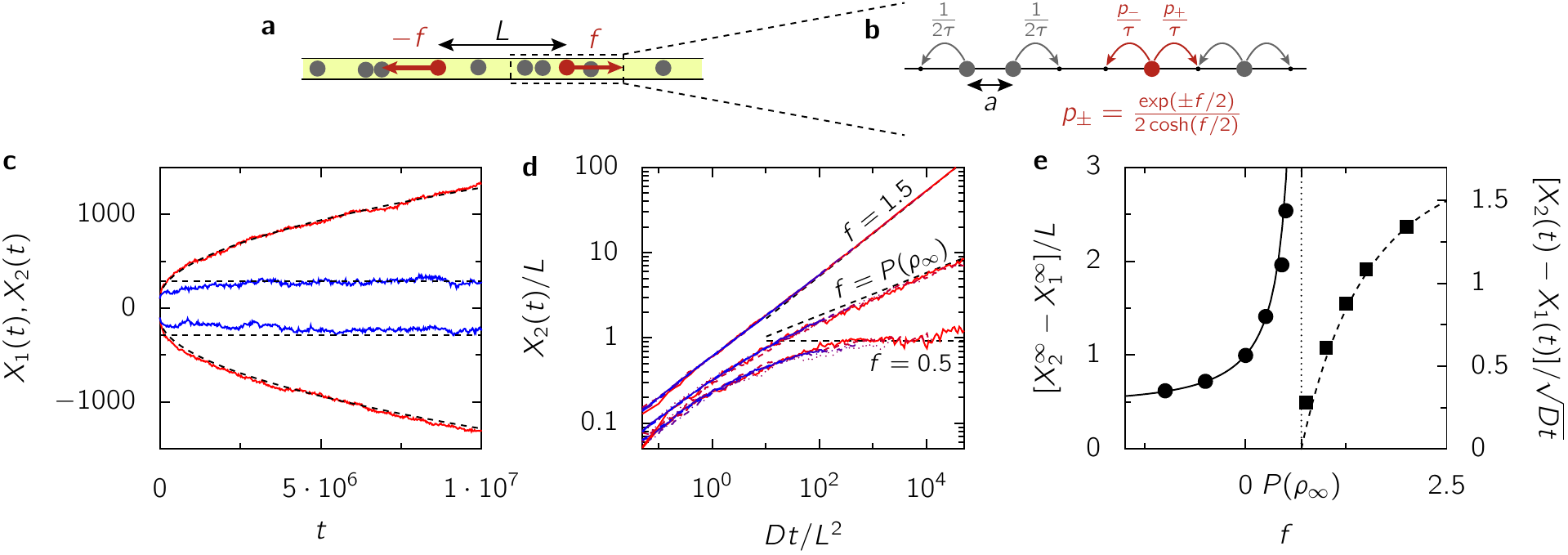}
\end{center}
\caption{\textbf{Two probes submitted to opposite forces in a single file system.}
{\bf a}, General scheme: two probes (red) in a single file system (bath particles in grey) initially at a distance $L$ are submitted to opposite forces $\mp f$.
{\bf b}, Possible moves and transition rates in the SEP with biased probes (red).
{\bf c}, Example of trajectories for $f=0.5$ (blue) and $f=1.5$ (red), for $\rho_\infty=0.5$. The dashed black lines are the theoretical predictions.
Distances are given in units of $a$ and forces in units of $\kT/a$.
{\bf d}, Rescaled trajectory of the probe 2 in log-log scale for $f=0.5,\, P(\rho_\infty)\simeq 0.69,\,1.5$, and $L=10, 20, 50, 100, 200, 500$ (red to blue).
{\bf e}, Separation of the probes in the bound ($f<P(\rho_\infty)$, $\CIRCLE$) and unbound ($f>P(\rho_\infty)$, $\blacksquare$) regimes. 
Points are the results of numerical simulations and the lines are the theoretical results (\ref{eq:bound_asympt}) and (\ref{eq:unbound_asympt}).}
\label{fig:asym}
\end{figure*}

\begin{figure*}
\begin{center}
\includegraphics[scale=.9]{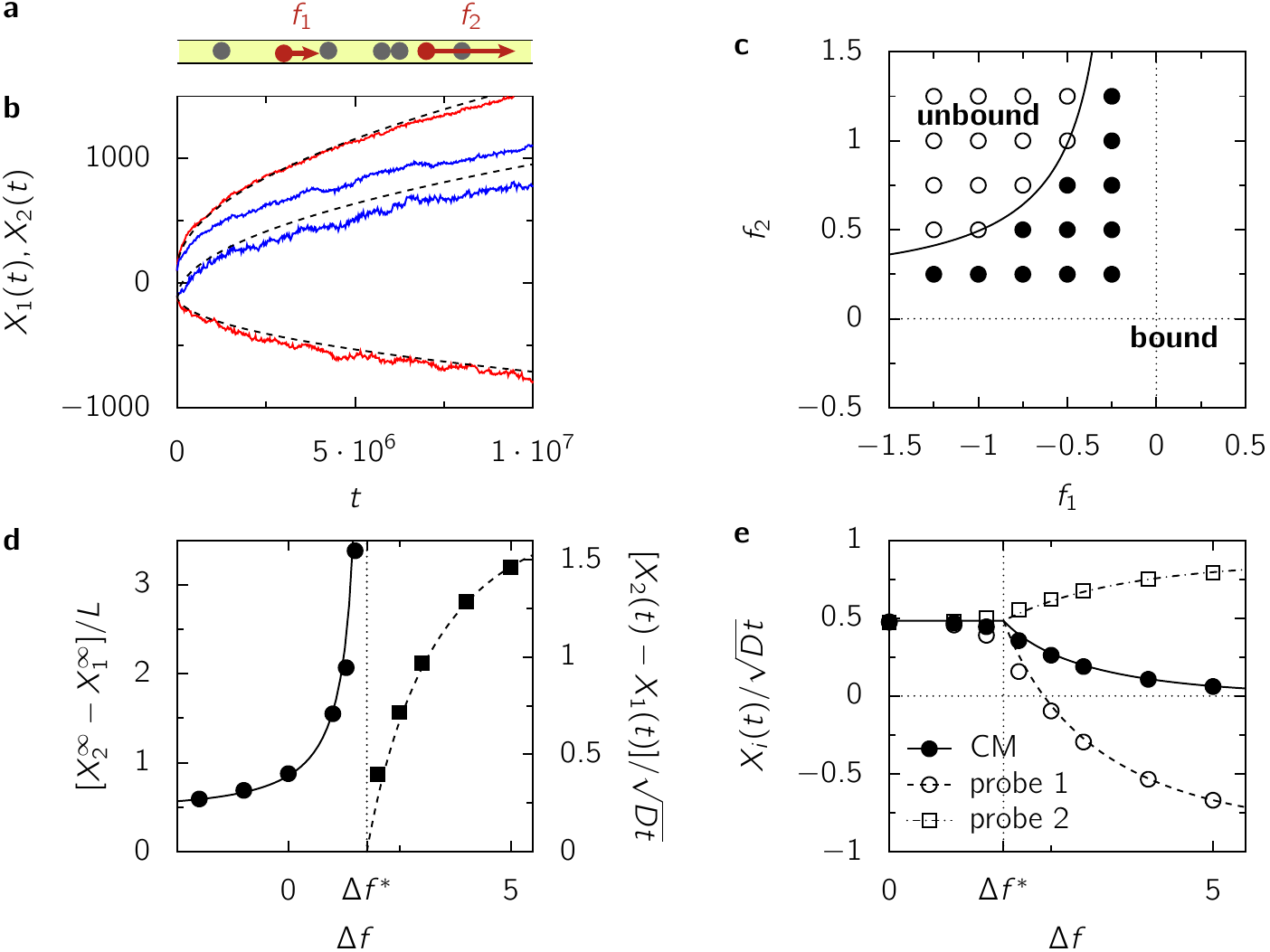}
\end{center}
\caption{
\textbf{Two probes submitted to arbitrary forces.}
{\bf a}, Two probes located at $X_1<X_2$ are submitted to arbitrary forces $f_1$ and $f_2$.
{\bf b}, Two examples of trajectories for $f_1=0$, $f_2=1$ (blue) and $f_1=-1$, $f_2=2$ (red).
{\bf c}, Phase diagram: bound ($\CIRCLE$) and unbound ($\Circle$) configurations, the line is the theoretical prediction (\ref{eq:f1c},\ref{eq:f2c}).
{\bf d}, Separation in the bound ($\CIRCLE$) and unbound ($\blacksquare$) regimes for $F=f_1+f_2=1$ as a function of the force difference $\Delta f=f_2-f_1$; $\Delta f^*$ is the critical force difference (Eqs.~(\ref{eq:f1c},\ref{eq:f2c})).
{\bf e}, Motion of each probe, and of the center of mass (CM) for the same parameters.
}
\label{fig:F}
\end{figure*}

\begin{figure*}
\begin{center}
\includegraphics[scale=.9]{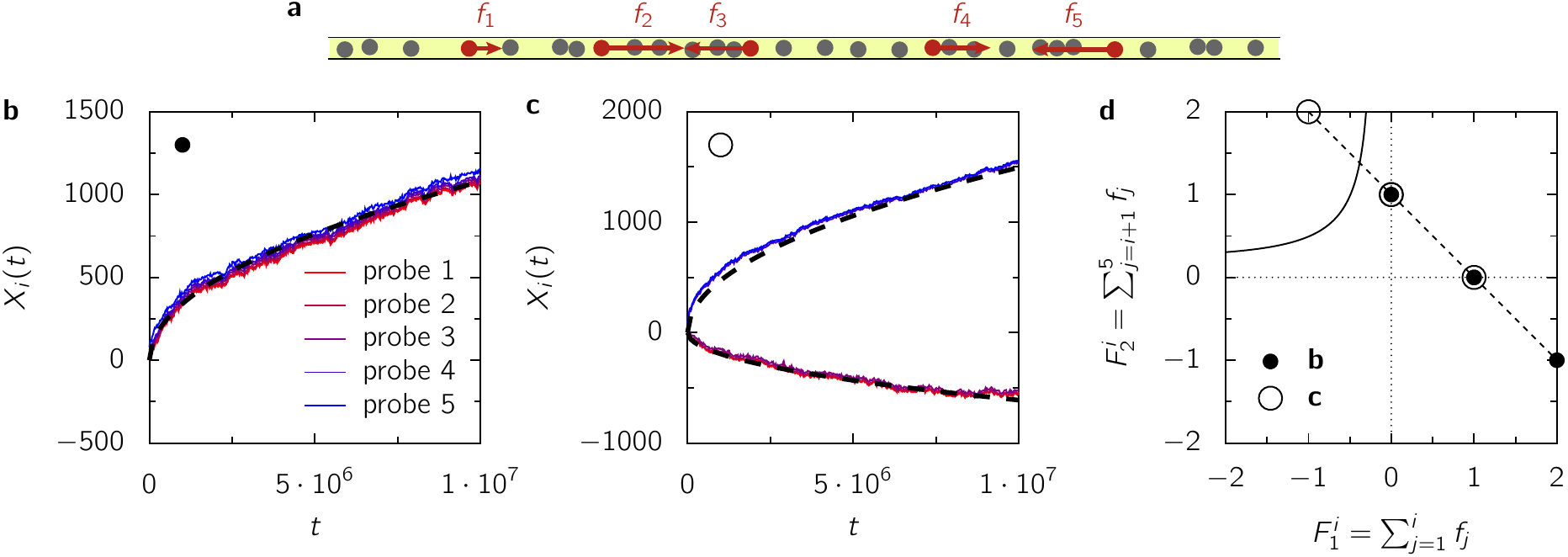}
\end{center}
\caption{
\textbf{Five probes submitted to arbitrary forces (a).}
{\bf b}, {\bf c}, Simulated trajectories and theoretical predictions (black dashed lines) for forces $(1, 1, -1, -1, 1)$ ({\bf b}) and $(1, -1, -1, 1, 1)$ ({\bf c}).
{\bf d}, Two probes phase diagram for $\rho=0.5$; it shows that all the possible divisions remain bound in case {\bf b} ($\CIRCLE$) and that the groups $(1,2,3)$ and $(4,5)$ unbind in case {\bf c} ($\Circle$).}
\label{fig:5probes}
\end{figure*}

Here, we determine the response of several probes to external forces (Figs.~\ref{fig:asym}a, \ref{fig:F}a, \ref{fig:5probes}a).
We show that the bonds induced by the single-file geometry can be broken; 
we characterize this unbinding transition, and explain its impact on the motion of the probes.
We obtain exact results for the average positions of the probes in the simple exclusion process (SEP), which is a paradigmatic model of single-file systems.
These conclusions are shown to also apply to model colloidal systems used in experiments~\cite{Wei2000, Lin2005}, which points towards their universality.

In the SEP, particles move on a one-dimensional lattice with step $a$; single-file diffusion is enforced by allowing at most one particle per site (Fig.~\ref{fig:asym}b).
The density $\rho$ is the proportion of occupied sites.
Each particle can jump to the left or to the right, with rates $1/(2\tau)$. 
For the probes, these rates are modified:
a probe submitted to an external force $f$ jumps to the left and to the right with rates $(1-s)/(2\tau)$ and $(1+s)/(2\tau)$, respectively, where $s=\tanh(af/[2\kT])$ is set by detailed balance.
Note that the gas of pointlike Brownian particles at density $\hat \rho$ is recovered as the limit of the SEP at vanishing density, $\rho\to 0$, with $\hat\rho=\rho/a$ kept constant.

First, we focus on the asymmetric case with two probes (Fig.~\ref{fig:asym}a).
Initially located at $X_1(0)=-L/2$ and $X_2(0)=L/2$, with $L\gg a$, with a uniform density $\rho_\infty$ of unbiased particles, they are submitted to forces $f_1=-f_2=-f$.
We performed numerical simulations~(App.~\ref{ap:sim_sep}) and observed two behaviors: they can either remain bound, or unbind and move away from each other, their displacement being proportionnal to $\sqrt{t}$ (Fig.~\ref{fig:asym}c,d).
In the bound state, the equilibrium distance between the probes increases with the force and diverges upon approaching a critical force; conversely, the factor of $\sqrt{t}$ in the unbound state decays to zero as the critical force is approached from above (Fig.~\ref{fig:asym}e).
At the critical force, the probes separate with a different exponent (Fig.~\ref{fig:asym}d).

Two probes submitted to external forces $f_1$ and $f_2$ can also be bound, and move together as $\sqrt{t}$, or unbound, and move as $A_i\sqrt{t}$ with $A_1\neq A_2$ (Fig.~\ref{fig:F}a,c).
Their state can be represented in a phase diagram (Fig.~\ref{fig:F}b).
Upon approaching the unbinding transition from above and below, the same behavior as in the antisymmetric case is found (Fig.~\ref{fig:F}d).
Interestingly, in the bound state the ``velocity'' of the probes depends only on the sum of the forces, $f_1+f_2$; when they unbind, the velocity of the center of mass decreases rapidly (Fig:~\ref{fig:F}e).
Finally, $N$ driven probes can also be bound and move as a whole (Fig.~\ref{fig:5probes}c) or separate into two groups (Fig.~\ref{fig:5probes}e).

\begin{figure}
\begin{center}
\includegraphics[scale=.9]{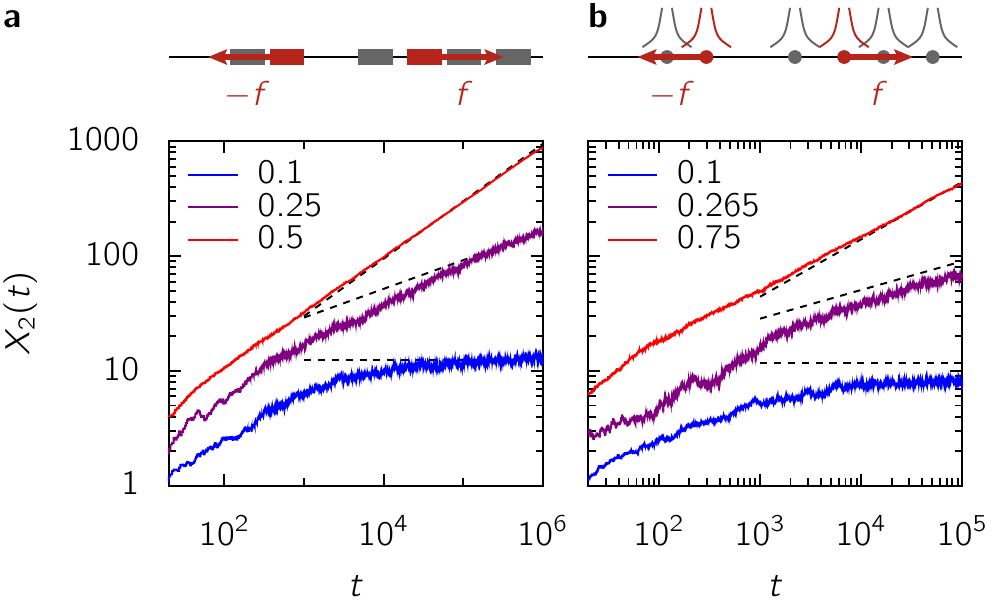}
\end{center}
\caption{{\bf Two probes submitted to opposite forces in the Tonks's gas (a) and the dipolar gas (b).} Trajectory of the probe 2 in the Tonks' gas of hard rods (\textbf{a}) and in the dipolar gas with $1/r^3$ interactions (\textbf{b}) at $\rho_\infty=0.2$, for different values of the force $f$ (solid lines). 
In the Tonks' gas, $P(\rho_\infty)=0.25$; in the dipolar gas, $P(\rho_\infty)\simeq 0.265$.
The dashed black lines are the theoretical predictions~(App.~\ref{ap:two_opp}).}
\label{fig:tonks_dipolar}
\end{figure}

We have run numerical simulations of other systems, focusing on the model systems used in experiments.
Systems implementing the single-file property have been realized with colloids confined to a narrow channel either printed in the substrate~\cite{Wei2000, Lin2005}, or generated with scanning optical tweezers~\cite{Lutz2004}.
The colloids either interact through a magnetic dipolar interaction, as $1/r^3$, where $r$ is the interparticle distance~\cite{Wei2000}, or behave as hard rods, as in the Tonks' gas~\cite{Lin2005}.
We simulated these two systems with an overdamped dynamics, inserting two probes submitted to opposite forces, and found the same phenomenology as in the SEP (Fig.~\ref{fig:tonks_dipolar}).

To account for these observations, we start from the hydrodynamic description of the SEP introduced in Refs.~\cite{Burlatsky1996, Landim1998} to investigate the response of a single probe to a constant force. 
Notably, this approach gives the exact result for the mean position of the probe and the density profile of the bath particles at long time.
The starting point of the analysis is that the bath density $\rho(x,t)$ has a diffusive behaviour~\cite{Spohn1991},
\begin{equation}
\frac{\partial \rho}{\partial t}(x,t)=D \frac{\partial^2\rho}{\partial x^2}(x,t),
\end{equation}
where the diffusion coefficient is $D=a^2/(2\tau)$.
The probe $i$, located at $X_i(t)$ in average, acts as a moving wall that imposes a no-flux boundary condition, namely, $D\frac{\partial \rho}{\partial x}(X_i^\pm,t) = -\rho(X_i^\pm,t) \frac{\dd X_i}{\dd t}$.

The several probes situation is conveniently analysed by first revisiting the single probe case.
Within the hydrodynamic approach, it has been shown~\cite{Burlatsky1996, Landim1998} that the densities immediately left and right of a probe moving as 
\begin{equation}\label{eq:x_ansatz}
X(t)\sim A\sqrt{t}
\end{equation}
are given by
\begin{equation}\label{eq:rho_pm}
\rho(X^\pm) = \rho_\infty g\left(\pm \frac{A}{2\sqrt{D}}\right),
\end{equation}
where $g(u)=[1-\sqrt{\pi}u\exp(u^2)\erfc(u)]^{-1}$.
The system of equations is closed with a relation between the velocity of the probe, the force on the probe and the densities on each side of the probe~\cite{Burlatsky1996,Landim1998}. 
We show in App.~\ref{ap:single} that this relation can actually be interpreted as a force balance,
\begin{align}\label{eq:force_bal}
f = P(\rho(X^+))-P(\rho(X^-)),
\end{align}
which involves the pressure of the SEP~\cite{Hill1960},
\begin{equation}\label{eq:eos}
P(\rho)=-\frac{\kT}{a}\log(1-\rho).
\end{equation}
Using Eqs.~(\ref{eq:rho_pm},\ref{eq:force_bal}) gives back the implicit equation for $A$ given in Refs.~\cite{Burlatsky1996, Landim1998}, which can be solved numerically.
As we proceed to show, this new interpretation allows a direct generalization to the case of several driven particles.
Moreover, it underlines the robustness of our approach, which can be applied to other single-file systems.

We turn to the situation where two probes are submitted to opposite external forces, $f_2=-f_1=f$ (Fig.~\ref{fig:asym}).
First, we focus on the case where the probes remain bound, meaning that their positions converge, and we define $X_i^\infty=\lim_{t\to\infty}X_i(t)$ (Fig.~\ref{fig:asym}c,d).
In this case, the density between the probes is uniform and we denote it by $\rho_1$, while the density outside of the probes is the density at infinity, $\rho_\infty$.
The density between the probes is given by Eq.~(\ref{eq:force_bal}), $P(\rho_1)=P(\rho_\infty)-f$, and allows one to compute the equilibrium distance between the probes, $L^\infty=X_2^\infty-X_1^\infty=L\rho_\infty/\rho_1$ (Fig.~\ref{fig:asym}d).
This bound state is observed as long as the force $f$ does not exceed the pressure of the outer gas, $P(\rho_\infty)$.
As this pressure is approached from below, the distance between the probes diverges as (Fig.~\ref{fig:asym}e)
\begin{equation}\label{eq:bound_asympt}
\frac{X_2^\infty-X_1^\infty}{L}\underset{f\to P(\rho_\infty)^-}{\sim} 
\frac{\rho_\infty P'(0)}{P(\rho_\infty)-f}
\stackrel{\mathrm{SEP}}{=}\frac{\kT \rho_\infty}{a[P(\rho_\infty)-f]},
\end{equation}
where $P'(\rho)$ denotes the derivative of the pressure with respect to $\rho$.

When the forces overcome the pressure of the gas, the probes unbind and move apart as $X_2(t)=-X_1(t)\sim A\sqrt{t}$, and the density between the probes decays to zero. 
The force balance (\ref{eq:force_bal}) for the probe 2 together with Eq.~(\ref{eq:rho_pm}) give $f=P(\rho_\infty g(A/[2\sqrt{D}]))$, which is an implicit equation for $A$~(Fig.~\ref{fig:asym}d).
As $f$ approaches $P(\rho_\infty)$ from above, $A$ decays and (Fig.~\ref{fig:asym}e)
\begin{align}
\frac{X_2(t)}{\sqrt{D t}} &\underset{f\to P(\rho_\infty)^+}{\underset{t\to\infty}{\sim}} \frac{2}{\sqrt{\pi}}\frac{f-P(\rho_\infty)}{\rho_\infty P'(\rho_\infty)} \label{eq:unbound_asympt}\\
& \quad \stackrel{\mathrm{SEP}}{=} \frac{2}{\sqrt{\pi}}\frac{1-\rho_\infty}{\rho_\infty}\frac{a[f-P(\rho_\infty)]}{\kT}.
\end{align}

Eqs.~(\ref{eq:bound_asympt}-\ref{eq:unbound_asympt}) quantify the behaviour of the system at the vicinity of the unbinding transition, which occurs at $f=P(\rho_\infty)$.
However, they leave aside the important question of what happens at the transition.
From Eqs.~(\ref{eq:bound_asympt}-\ref{eq:unbound_asympt}), we may expect the separation to evolve in time as a power law, $X_2(t)=-X_1(t)\sim Ct^\gamma$, with a different exponent $\gamma\in(0,1/2)$.
Under this assumption, the density $\rho_1(t)$ between the two probes is uniform and $\rho_1(t)\sim L\rho_\infty/(2Ct^\gamma)$.
The density in front of the probe 2 can be shown to be given by $\rho(X_2(t)^+,t)-\rho_\infty\propto \rho_\infty C t^{\gamma-\frac{1}{2}}$.
Using Eqs.~(\ref{eq:force_bal},\ref{eq:eos}) leads to $X_2(t)\propto\sqrt{LP'(0)/P'(\rho_\infty)}t^{1/4}$ (App.~\ref{ap:two_opp}), and the exact expression is
\begin{align}
X_2(t) & \underset{t\to\infty}{\sim} \sqrt{\frac{2\sqrt{\pi}}{\mathrm{B}(1/2,1/4)}}\sqrt{\frac{P'(0) L}{P'(\rho_\infty)}} (Dt)^{1/4}\\
& \stackrel{\mathrm{SEP}}{\simeq} 0.82\sqrt{(1-\rho_\infty)L} (Dt)^{1/4},
\end{align}
where $\mathrm{B}$ is the beta function (Fig.~\ref{fig:asym}d).

It is noteworthy that the dependence of the separation between the probes on the time $t$ and the initial separation $L$ is constrained by the diffusive scaling of the bath in the three regimes.
Indeed, the position of the second probe can be written in all regimes as
\begin{equation}
X_2(t) = L\psi(Dt/L^2),
\end{equation}
with $\psi(u)\sim 1$ if $f<P(\rho_\infty)$, $\psi(u)\sim u^{1/4}$ if $f=P(\rho_\infty)$ and $\psi(u)\sim \sqrt{u}$ if $f>P(\rho_\infty)$ (Fig.~\ref{fig:asym}d).

The considerations above can be extended to the case where the two probes are submitted to arbitrary forces $f_1$ and $f_2$ (Fig.~\ref{fig:F}a).
When the probes are bound, the density between them becomes uniform and the force balance (\ref{eq:force_bal}) shows that their displacement is $A\sqrt{t}$, where $A$ is the same as for a single probe submitted to the force $F=f_1+f_2$ (Fig.~\ref{fig:F}c,e, App.~\ref{ap:two_arb}).
Unbinding occurs when the forces overcome the pressure of the gas at the left of probe 1 and at the right of probe 2, i.e. when $f_1=f_1^*(A)$ and $f_2=f_2^*(A)$ (Fig.~\ref{fig:F}b) with
\begin{align}
f_1^*(A) & = -P\left(\rho_\infty g\left(-\frac{A}{2\sqrt{D}}\right)\right),\label{eq:f1c}\\
f_2^*(A) & = P\left(\rho_\infty g\left(\frac{A}{2\sqrt{D}}\right)\right).\label{eq:f2c}
\end{align}
After unbinding, the probes move as $X_i(t)\sim A_i\sqrt{t}$, $A_1<A_2$, with $f_1=-P(\rho_\infty g(-A_1/[2\sqrt{D}]))$ and $f_2=P(\rho_\infty g(A_2/[2\sqrt{D}]))$ (Fig.~\ref{fig:F}d,e).
The displacement of the center of mass does not depend on $\Delta f=f_2-f_1$ as long as the probes are bound, but it decreases rapidly when they unbind (Fig.~\ref{fig:F}e).

Our results show that two probes that are bound can be seen as a single one, and this statement directly generalizes to $N$ probes (Fig.~\ref{fig:5probes}a).
Moreover, when the ensemble of $N$ probes separates into two groups moving away from each other, each group can be seen as a single probe.
It is actually not possible to have more than two groups, except if there is a group of probes on which the total force is zero: a probe located between two separating probes sees a bath of vanishing density, and thus moves freely in the direction of its force, until it meets the left or right probe.
To determine whether the $N$ probes remain bound, the two probes analysis can be applied to the $N-1$ possible divisions of the $N$ probes into two groups.
The set of forces to consider are $(F_1^i, F_2^i)$, where $F_1^i=\sum_{j=1}^i f_j$ and $F_2^j=\sum_{j=i+1}^N f_j$, for $1\leq i<N$.
If all the points $(F_1^i, F_2^i)$ are in the bound region of the phase diagram in Fig.~\ref{fig:F}b, the $N$ probes remain bound (Fig.~\ref{fig:5probes}b,c), otherwise they split for the index $i$ that maximizes $\Delta F^i=F_2^i-F_1^i$ (Fig.~\ref{fig:5probes}d,e).

Our analytical results are in excellent agreement with numerical simulations. 
In fact, our results are expected to be exact because: 
(i) The hydrodynamic approach that we used has been shown to give exact results for the mean position of a single probe under a constant force at long times~\cite{Burlatsky1996,Landim1998}.
(ii) The motion of several probes that are bound can be computed exactly when the density is close to 1 using an expansion in the number of vacancies similar to the one used in Ref.~\cite{Benichou2013e}, and confirms our results.

We have provided exact results for the SEP, and have shown that the unbinding transition is robust, as it also takes place in continuous models that represent experimental systems~\cite{Wei2000, Lin2005}. 
Thus, the unbinding transition should be observed if driven particles are inserted in these systems, for instance dielectric colloids manipulated with a laser beam to simulate an external force~\cite{Berut2014, Martinez2017}.
Motile particles can also simulate an external force; for example, a few colloidal rollers, which are used as a model active matter system~\cite{Bricard2013}, could be incorporated in narrow channels with passive colloids.
At a larger scale, a mixture of active and passive vibrated disks can be confined to a circular channel~\cite{Briand2016, Junot2017, Briand2017}.

\begin{acknowledgments}
The work of O. B. is supported by the European Research
Council (Grant No. FPTOpt-277998).
We acknowledge discussions with D. Bartolo and S. Ciliberto about the possible experimental tests of our theoretical results.
\end{acknowledgments}

\appendix

\section{Numerical simulations of the Simple Exclusion Process}\label{ap:sim_sep}

\subsection{Details of the simulations}

$M$ particles are placed on a discrete line of size $N$ as follow: the positions of the probes (one to five)
are fixed deterministically and the positions of the others particles are assigned uniformly at random on the remaining sites.
At each iteration, a particle is chosen uniformly at random and an increment of time is drawn according to an exponential
law of parameter $N/\tau$ (this corresponds to the minimum of $N$ independant exponential laws of parameter
$1/\tau$). Then, the chosen particle jumps either to the left or to the right according to its given probabilities
(uniform for a bath particle, biased for a probe) if the neighboring site is not occupied.
Periodic boundary conditions are enforced.

In Figs. \ref{fig:asym}, \ref{fig:F} and \ref{fig:5probes} we used the following parameters:
$N=10\,000$, $\rho = 0.5$ ($M = 5000$), $\tau = 1$ and a final time $T = 10^7$.
We recorded the positions of the probes every $\Delta t = 10$.
Figs. \ref{fig:asym}c, \ref{fig:F}c, \ref{fig:5probes}c and \ref{fig:5probes}e show the linear evolution
of the positions for a single simulation. The other figures
(\ref{fig:asym}d, \ref{fig:asym}e, \ref{fig:F}b, \ref{fig:F}d, \ref{fig:F}e) correspond to an average over
40 to 100 simulations with the same parameters.

\subsection{Finite size effects}
\begin{figure*}
\begin{center}
\includegraphics[width=\linewidth]{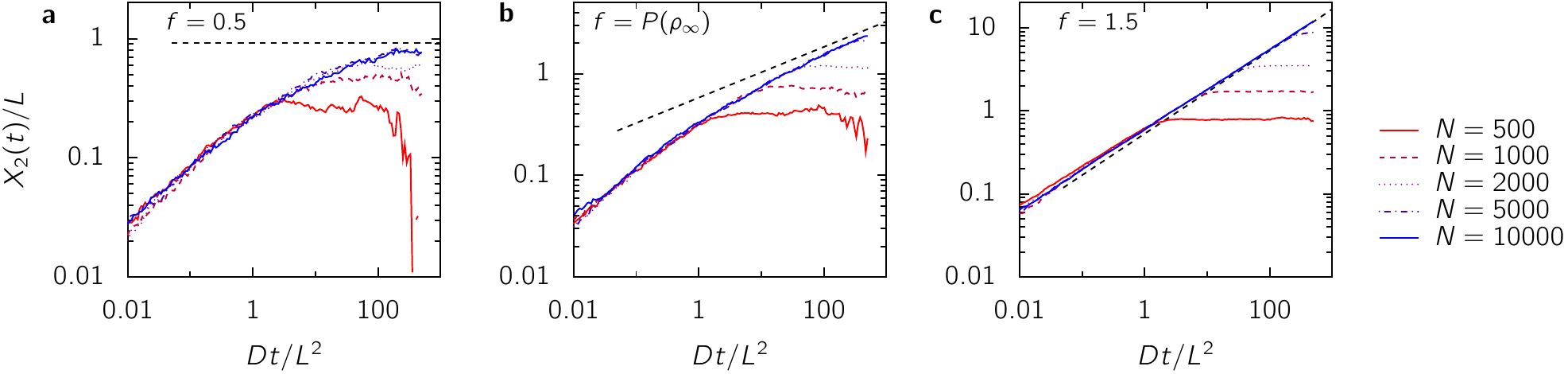}
\end{center}
\caption{Finite size effects for curves of Fig \ref{fig:asym}d. We chose $\rho_\infty = 0.5$, $L = 100$
and the number of particles ranges
from 500 to 10\,000. The final time is $T = 10^7$.
\textbf{a}, Below the critical force ($f=0.5$). \textbf{b}, at the critical force ($f=P(\rho_\infty)$).
\textbf{c}, above the critical force
($f=1.5$). The dashed black line is the asymptotic prediction.
}
\label{fig:finiteSize}
\end{figure*}

As we are conducting simulations on a finite line with periodic boundary conditions and comparing them against
predictions for the infinite line, we have to make sure that there are no finite size effects.
Fig. \ref{fig:asym}d is our most general figure and we focus on it:
in Fig. \ref{fig:finiteSize} we investigate the evolution of the
evolution of the curves in the three regimes as the number of particles is increased.
The larger the number of particles, the latter the positions saturate, and we conclude that $N=10\,000$
is indeed a good choice to avoid finite size effects (up to $T=10^7$).

\subsection{Phase diagram}\label{}
Fig. \ref{fig:F}b shows the phase diagram at density $\rho_\infty = 0.5$.
Given data for $X_1(t), X_2(t)$ up to $T=10^7$, with forces $f_1$ and $f_2$, we need a criterion to determine
whether the probes are in the bound or unbound state. In the bound state we expect $X_2 - X_1 \propto t^0$ while in
the unbound state $X_2 - X_1 \propto t^{1/2}$ (remind that the critical state gives $t^{1/4}$).
In the range $t\in [T/4, T]$ (to get rid of the transitory regime), we fit $\log (X_2-X_1)$ versus $\log (t)$
as a line. The slope gives the exponent of $X_2 - X_1$ as a power law in $t$.
If this numerical exponent is lower than $0.25$, we classify the system as bound, else as unbound.

\begin{figure*}
\begin{center}
\includegraphics[]{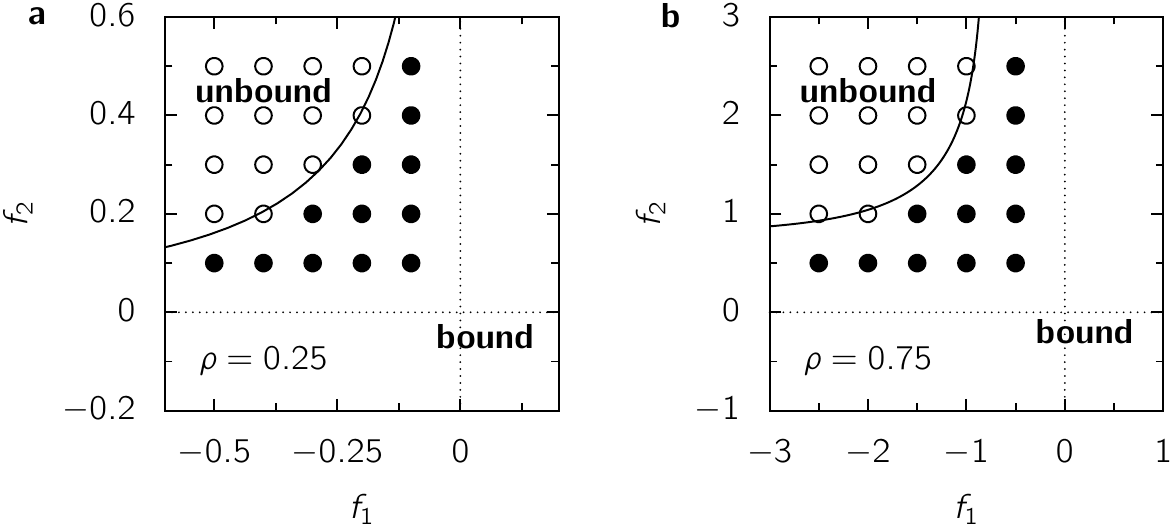}
\end{center}
\caption{
Phase diagram: bound ($\CIRCLE$) and unbound ($\Circle$) configurations, the line is the theoretical prediction.
\textbf{a}, density $\rho_\infty = 0.25$. \textbf{b}, density $\rho_\infty = 0.75$.
}
\label{fig:phaseDiag}
\end{figure*}

In this SI, we provide two additional phase diagrams (Fig. \ref{fig:phaseDiag}) at densities $\rho_\infty = 0.25$ and $0.75$.
The behavior observed from the simulations is in agreement with the theoretical prediction.

\section{Numerical simulations of continuous systems}\label{ap:sim_cont}

A simulation of a continuous system with $N$ particles at density $\rho$ starts by
placing uniformly at random the particles on a line of size $L_0 = N/\rho$.
Periodic boundary conditions are enforced. 
The particles
follow a discretized Langevin equation: given a timestep $\Delta t$ the position
$x_i$ of particle $i$ evolves according to the following equation:
\begin{equation}
 x_i(t+\Delta t) = x_i(t) + F_i \Delta t + \sum_{j\neq i} f_{j\to i} \Delta t
 + \Upsilon_{i,t} \sqrt{2T\Delta t}
\end{equation}
\begin{itemize}
 \item $F_i$ is the external force applied on particle $i$ (it is zero for the bath
 particles, which are not biased).
 \item $T$ is the temperature and is always set to $T=1$.
 \item $\Upsilon_{i,t}$ is a random number generated according to a standard normal distribution.
 \item $f_{j\to i}$ is the force from particle $j$ on particle $i$.
 \begin{itemize}
  \item In the case of the Tonks gas, we consider interactions between nearest neighbors
  according to one-sided springs.
  \begin{equation}
  f_{i-1\to i} = \epsilon \left[\sigma - (x_i - x_{i-1})\right]
  \Theta\left(a - [x_i - x_{i-1}]\right)
  \end{equation}
  $\Theta$ is the Heaviside step function. $a = 1$ is the length of a rod.
  $\epsilon$ is the strength of the potential, we chose $\epsilon = 100$ so that
  the rods are close to hard rods.
  \item In the case of the dipole-dipole interaction, we consider a potential
  $V(r) = A/r^3$. To be close to the experiments \cite{Wei2000}, we place the particles
  on a circle of radius $R = L_0/(2\pi)$: the energy $E_{ij}$ of interaction between
  particles $i$ and $j$ is
  \begin{equation}
   E_{ij} = \frac{A}{R^3 (2[1-\cos(\theta_{ij})])^{3/2}}
  \end{equation}
  with $\theta_{ij} = 2\pi(x_i - x_j)/L_0$ (in periodic boundary conditions).
  The force is
  \begin{equation}
   f_{j\to i} = \frac{6\sqrt{2}\pi^4 A}{L_0^4} \frac{\sin(\theta_{ij})}{[1-\cos(\theta_{ij})]^{5/2}}
  \end{equation}
 \end{itemize}
\end{itemize}

We check at each iteration that the particles do not cross. 
If such an event occurs, we restart the simulation.

When two probes are considered, we impose their difference of indice $\Delta i$
(i.e., there are $\Delta i - 1$ particles between them)
so that the initial distance is on average $L=\Delta i / \rho$.
All the observables are averaged over multiple simulations (20 to 500).

For the simulations of the Tonks gas (Fig. \ref{fig:tonks_dipolar}),
the following parameters were used:
$\rho = 0.2$, $\epsilon = 100$, $\Delta t = 0.005$ and $\Delta i = 10$.
The number of particles is $N = 500$ for $f = 0.1$ and $N = 1000$ for $f=0.25$ and $0.5$.
For the dipole-dipole interaction, the parameters are:
$\rho = 0.2$, $A = 1$, $\Delta t = 0.001$ and $\Delta i = 10$.
The number of particles is $N=200$ for $f=0.1$ and $N=400$ for $f=0.265$ and $0.75$.

\section{Governing equations}\label{ap:gov_eq}

Here we derive the governing equations for the position of the probe and the density field of the gas.

\subsection{Position of the probe}\label{}

The probe is submitted to a bias $s\in[-1,1]$: if its neighboring sites are empty, it jumps to the right at rate $(1+s)/(2\tau)$ and to the left at rate $(1-s)/(2\tau)$.
Detailed balance enforces a relation between the bias $s$ and the force $f$:
\begin{equation}
\frac{1+s}{1-s}=\exp \left(\frac{af}{\kT} \right).
\end{equation}
This relation also reads
\begin{equation}
s = \tanh \left(\frac{af}{2 \kT}  \right)
\end{equation}

In the lattice gas, the jumps can occur only if the final sites are empty; this is the case with probability $1-\rho(X^+)$ on the right, and probability $1-\rho(X^-)$ on the left.
Finally, the average move gives the velocity of the probe:
\begin{align}
V
& =\frac{\dd X}{\dd t}\\
& =V_0\left(\frac{\rho(X^-)-\rho(X^+)}{2}+s \left[1-\frac{\rho(X^+)+\rho(X^-)}{2} \right]\right), \label{eq:vel}
\end{align}
where
\begin{equation}
V_0=\frac{a}{\tau}.
\end{equation}

\subsection{Density field}\label{}

The density field has a diffusive dynamics~\cite{Spohn1991}:
\begin{equation}
\frac{\partial\rho}{\partial t}(x,t)=D \frac{\partial^2\rho}{\partial x^2}(x,t),
\end{equation}
with diffusion coefficient
\begin{equation}
D=\frac{a^2}{2\tau}.
\end{equation}
The diffusion coefficient can be computed for a single particle on the lattice: the variance of its position after a time $t$ is $\langle x_t^2 \rangle = a^2t/\tau = 2Dt$.
We can introduce the current $j(x)$:
\begin{align}
\frac{\partial\rho}{\partial t} & = -\frac{\partial j}{\partial x}\\
j & = -D \frac{\partial \rho}{\partial x}.
\end{align}

We can write the dynamics for the density field in the reference frame of the probe,
\begin{equation}
\rho^*(x,t)=\rho(x+X(t),t);
\end{equation}
we get
\begin{equation}
\frac{\partial\rho}{\partial t}^*=D \frac{\partial^2\rho^*}{\partial x^2}+V \frac{\partial \rho^*}{\partial x}=-\frac{\partial}{\partial x} \left(-D \frac{\partial\rho^*}{\partial x}-V\rho^* \right).
\end{equation}

\subsection{Boundary condition close to the probe}\label{}

For the gas, the probe is a hard wall; hence, in the reference frame of the probe, the current should vanish,
\begin{equation}
j^*(0^\pm)=-D \frac{\partial\rho^*}{\partial x}(0^\pm)-V\rho^*(0^\pm)=0,
\end{equation}
leading to
\begin{align}
D\frac{\partial\rho}{\partial x}(X^\pm) & = -V\rho(X^\pm),\\
D\frac{\partial\rho^*}{\partial x}(0^\pm) & =-V\rho^*(0^\pm).
\end{align}

\section{Single probe with a constant bias}\label{ap:single}

Here we derive the behavior of a single probe with a constant bias.
Our derivation is close to the one given in Ref.~[\citenum{Burlatsky1996}]; we reproduce it here and show how it can be reinterpreted in term of the pressure of the SEP.

\subsection{Equations}\label{}

We work in the reference frame of the probe.
The equations that we have to solve are
\begin{align}
\frac{V(t)}{V_0} & = \frac{\rho(0^-,t)-\rho(0^+,t)}{2}\nonumber\\
& \qquad +s \left[1-\frac{\rho(0^+,t)+\rho(0^-,t)}{2} \right],\label{eq:1t_vel}\\
\frac{\partial\rho^*}{\partial t}(x,t) & = D \frac{\partial^2\rho^*}{\partial x^2}(x,t)+V(t)\frac{\partial\rho^*}{\partial x}(x,t),\label{eq:1t_density}\\
D \frac{\partial\rho^*}{\partial x}(0^\pm,t) & = -V(t)\rho^*(0^\pm,t).\label{eq:1t_bct}
\end{align}
The initial and boundary conditions are
\begin{align}
\rho(x,0) & = \rho_\infty,\\
\rho(\pm\infty,t) & = \rho_\infty.
\end{align}

\subsection{Solution using a diffusive scaling}\label{sub:sol_diff}

Since the density in the reference frame of the probe follows a diffusion equation with a bias, we may expect a diffusive scaling for the solution:
\begin{equation}
\rho^*(x,t)=\phi \left(\frac{x}{\sqrt{D t}} \right).
\end{equation}
We show later that there is no need for a time dependent factor.

In Eq.~(\ref{eq:1t_density}), this leads to
\begin{equation}
-\frac{x}{2\sqrt{D}t^{3/2}}\phi'\left(\frac{x}{\sqrt{D t}} \right)=\frac{1}{t}\phi'' \left(\frac{x}{\sqrt{D t}} \right)+\frac{V(t)}{\sqrt{D t}}\phi'\left(\frac{x}{\sqrt{D t}} \right).
\end{equation}
The diffusive scaling holds if 
\begin{equation}
V(t)=\frac{\tA\sqrt{D}}{\sqrt{t}};
\end{equation}
we use this ansatz from now on. 
Note that $\tA$ is related to the constant $A$ used in the main text (equation (\ref{eq:x_ansatz})) through $A=2\sqrt{D}\tA$.
The equation for $\phi$ reads now
\begin{equation}
\phi''(u)=-\left(\frac{u}{2}+\tA \right)\phi'(u).
\end{equation}
Its solution is
\begin{equation}
\phi'(u)=\phi'(0)\ed^{-\frac{u^2}{4}-\tA u}.
\end{equation}
We can deduce the density profile as a function of $\phi'(0)$: for $u>0$, 
\begin{align}
\phi(u) & = \phi(\infty)-\int_u^\infty \phi'(u')\dd u'\\
& = \rho_\infty-\phi'(0^+)\int_u^\infty\ed^{-\frac{u'^2}{4}-\tA u'}\dd u'\\
& = \rho_\infty-\phi'(0^+)\ed^{\tA^2}\int_u^\infty\ed^{-(u'+\tA)^2/4}\dd u'\\
& = \rho_\infty-\phi'(0^+)\ed^{\tA^2}2\int_{\frac{u}{2}+\tA}^\infty\ed^{-w^2}\dd w\\
& = \rho_\infty-\phi'(0^+)\ed^{\tA^2}\sqrt{\pi}\erfc \left(\frac{u}{2}+\tA\right).\label{eq:1t_profile_pos}
\end{align}
For $u<0$ we get
\begin{equation}
\phi(u) = \rho_\infty+\phi'(0^-)\ed^{\tA^2}\sqrt{\pi}\erfc \left(-\frac{u}{2}-\tA\right).
\end{equation}

Now, we use Eq.~(\ref{eq:1t_bct}) to relate $\phi'(0)$ to $\phi(0)$: 
\begin{equation}
\phi'(0^\pm)=-\tA\phi(0^\pm).
\end{equation}
In Eq.~(\ref{eq:1t_profile_pos}), this leads to
\begin{equation}
\phi(0^+)=\rho_\infty+\tA\phi(0^+)\ed^{\tA^2}\sqrt{\pi}\erfc (\tA),
\end{equation}
hence 
\begin{equation}\label{eq:phi_0p}
\phi(0^+)=\rho_\infty g(\tA),
\end{equation}
with
\begin{equation}
g(\tA) = \frac{1}{1-\sqrt{\pi}\tA\ed^{\tA^2}\erfc (\tA)}
\end{equation}
We also get
\begin{equation}\label{eq:phi_0m}
\phi(0^-)=\rho_\infty g(-\tA).
\end{equation}

In equation~(\ref{eq:1t_vel}), the left hand side term decays to zero at long times, and the two terms on the right hand side should cancel, leading to
\begin{equation}
\rho_\infty[g(\tA)-g(-\tA)] = s(2-\rho_\infty[g(\tA)+g(-\tA)]),
\end{equation}
where we have used equations (\ref{eq:phi_0p},\ref{eq:phi_0m}).
This relation can be written as
\begin{equation}\label{eq:implicit_A}
\frac{1-\rho_\infty g(\tA)}{1-\rho_\infty g(-\tA)}=\frac{1-s}{1+s}=\exp \left(-\frac{af}{\kT} \right).
\end{equation}
This relation is an implicit equation for $\tA$, which should be solved numerically.

\subsection{Interpretation with the pressure of the SEP}\label{}

The pressure of the SEP is given by~\cite{Hill1960}
\begin{equation}\label{eq:eos_si}
P(\rho)=-\frac{\kT}{a}\log(1-\rho).
\end{equation}
This expression allows to rewrite equation~(\ref{eq:implicit_A}) as
\begin{equation}\label{eq:force_bal_si}
P(\rho(X^+))-P(\rho(X^-))=f,
\end{equation}
where $\rho(X^\pm)=\rho_\infty g(\pm \tA)$ is the density in front of or behind the probe.
which is the force balance (\ref{eq:force_bal}) given in the main text.

\subsection{Analytical result at small force or high density}\label{}

Analytical results can be obtained when $\tA\ll 1$, which corresponds to small force or high density.
First, we can use the expansion of $g$ around 0:
\begin{equation}
g(\tA)\simeq 1+\sqrt{\pi} \tA.
\end{equation}
In equation (\ref{eq:force_bal_si}), this gives
\begin{equation}
2\sqrt{\pi}\rho_\infty P'(\rho_\infty) \tA = f,
\end{equation}
where $P'(\rho)$ denotes the derivative of $P$ with respect to $\rho$.
Using the definition of $\tA$, we get that the displacement is given by
\begin{equation}
X(t)
\underset{t\to\infty}{\sim} \frac{f}{\sqrt{\pi}\rho_\infty P'(\rho_\infty)}\sqrt{Dt}
\stackrel{\mathrm{SEP}}{=} \frac{(1-\rho_\infty)}{\sqrt{\pi}\rho_\infty}\frac{af}{\kT}\sqrt{Dt}
\end{equation}
where we have used the equation of state (\ref{eq:eos_si}) for the pressure of the SEP to get the second relation.

\section{Two probes submitted to opposite forces}\label{ap:two_opp}

We consider two probes submitted to opposite forces: $f_1=-f_2=-f$.

\subsection{Arrested configuration}\label{}

Here, we are interested in the arrested situation where the position of the probes converges to a constant value.
At long times, the density between the probes becomes uniform, and we denote it $\rho_1$.
The density outside the probes is also uniform, and equal to $\rho_\infty$.

Here, equation (\ref{eq:vel}) also reduces to the force balance (\ref{eq:force_bal_si}) at long times; writing it for the probe 2 leads to
\begin{equation}
P(\rho_\infty)-P(\rho_1)=f.
\end{equation}

As $f$ approaches $P(\rho_\infty)$ from below, $\rho_1\to 0$ so that we can expand, $P(\rho_1)\simeq P'(0)\rho_1$, hence
\begin{equation}
\rho_1\simeq \frac{P(\rho_\infty)-f}{P'(0)}.
\end{equation}
The final distance between the probes is related to the density through the conservation of the number of particles:
\begin{equation}
\frac{X_2^\infty-X_1^\infty}{L}=\frac{\rho_\infty}{\rho_1}\underset{f\to P(\rho_\infty)^-}{\sim} \frac{\rho_\infty P'(0)}{P(\rho_\infty)-f}.
\end{equation}

\subsection{Probes moving apart}\label{}

When $f>P(\rho_\infty)$, the probes move appart.
We still expect their velocity to scales as $V_1(t)\sim -V_2(t) \sim - \tA\sqrt{D/t}$ at long times, so that the density in front of the probe 2 is $\rho(X_2^+)=\rho_\infty g(\tA)$.

Equation (\ref{eq:vel}) reduces to the force balance (\ref{eq:force_bal_si}) at long times; writing it for the probe 2 leads to
\begin{equation}
P(\rho_\infty g(\tA))=f,
\end{equation}
which is an implicit equation for $\tA$.

As $f$ approaches $P(\rho_\infty)$ from above, $\tA$ approaches 0 so that we can expand
\begin{align}
P(\rho_\infty g(\tA)) &\simeq P(\rho_\infty)+\tA g'(0) P'(\rho_\infty) \\
& = P(\rho_\infty)+\sqrt{\pi} P'(\rho_\infty) \tA.
\end{align}
In the relation above, we thus get $\sqrt{\pi} P'(\rho_\infty) \tA\simeq f-P(\rho_\infty)$; finally, the displacements are
\begin{equation}
X_2(t)\underset{t\to\infty}{\sim} \frac{2}{\sqrt{\pi}} \frac{f-P(\rho_\infty)}{P'(\rho_\infty)}\sqrt{Dt}.
\end{equation}




\subsection{Critical regime}\label{sub:critical}

\subsubsection{Density in front of the probe for an arbitrary velocity}\label{}

We are interested in the behavior of the probes in the critical regime, i.e. when $f=P(\rho_\infty)$.
From the behavior of the system as the transition is approached from below and above, we may expect that the position of the probe 2 follows $X_2(t)\sim t^\gamma$, with $0<\gamma<1/2$.

In order to determine the behavior of the probes, we have to determine the density in front of a probe with an arbitrary time-dependent velocity $V(t)$.
In the reference frame of the probe, the density evolves according to equations~(\ref{eq:1t_density},\ref{eq:1t_bct}).

In equation~(\ref{eq:1t_density}), the two terms on the right hand side have the same order of magnitude if $V(t)\sim t^{-1/2}$, but we can expect the second term to be negligible if the velocity decays faster than that.
Another point of view is to say that at long times, the velocity $V(t)$ and density variations $\delta\rho^*$ are small, and that the second term is the product of two small terms. 
Applying the same reasoning to Eq.~(\ref{eq:1t_bct}), we keep
\begin{align}
\frac{\partial\rho^*}{\partial t}(x,t) & = D \frac{\partial^2\rho^*}{\partial x^2}(x,t),\\
D \frac{\partial \rho^*}{\partial x}(0^+,t) & = -V(t)\rho_\infty.
\end{align}

The solution to this equation reads
\begin{equation}
\delta\rho^*(x,t)=\rho^*(x,t)-\rho_\infty=\rho_\infty\int_0^t V(t')G(x,t-t')\dd t',
\end{equation}
where
\begin{equation}
G(x,t)=\frac{1}{\sqrt{\pi D t}}\ed^{-\frac{x^2}{4Dt}}.
\end{equation}
We deduce the density in front of the probe:
\begin{equation}
\delta\rho_+(t)=\delta\rho^*(X^+,t) = \frac{1}{\sqrt{ \pi D}}\rho_\infty \int_0^t \frac{V(t')}{\sqrt{t-t'}}\dd t'.
\end{equation}

Assume now that
\begin{equation}
V(t)=C\sqrt{D}t^{\gamma-1},
\end{equation}
then
\begin{align}
\delta\rho_+(t)
& = \frac{C\rho_\infty}{\sqrt{\pi}} \int_0^t \frac{t'^{\gamma-1}}{\sqrt{t-t'}}\dd t'\\
&_= \frac{C\rho_\infty}{\sqrt{\pi}} t^{\gamma-\frac{1}{2}}\int_0^1 \frac{u^{\gamma-1}}{\sqrt{1-u}}\dd u.
\end{align}
The integral, that we denote $b_\gamma$, is given by the beta function $\mathrm{B}$,
\begin{equation}
b_\gamma=\int_0^1 \frac{u^{\gamma-1}}{\sqrt{1-u}}\dd u = \mathrm{B} \left(\frac{1}{2},\gamma \right).
\end{equation}

If $\gamma=1/2$, the integral is $\pi$ and we get $\delta\rho_+ = \sqrt{\pi}\rho_\infty C$, which is the exact result (\ref{eq:phi_0p}) in the limit $C\ll 1$.

\subsubsection{Displacement of the probe}\label{}

We start with a scaling law argument.
We assume that the position of the probe 2 follows $X(t)\sim t^\gamma$.
Then the density between the probes decays as $\rho_-(t)\sim 1/X(t)\sim t^{-\gamma}$.
The density in front of the probe 2 follows $\delta\rho_+(t)\sim t^{\gamma-\frac{1}{2}}$.
Balancing these two terms in Eq.~(\ref{eq:vel}) leads to $-\gamma=\gamma-\frac{1}{2}$, and thus to $\gamma=1/4$.
In this case, the velocity decays as $V(t)\sim t^{-3/4}$ and thus does not contribute in Eq.~(\ref{eq:vel}): the force balance (\ref{eq:force_bal_si}) still applies.

We now assume that $X(t)= C t^{1/4}$, $V(t)= \frac{C}{4}t^{-3/4}$.
The complete expressions for the density between the probes, $\rho_-(t)$, and in front of the probe 2, $\rho_+(t)=\rho_\infty+\delta\rho_+(t)$ are now
\begin{align}
\rho_-(t) &  = \rho_\infty\frac{L}{2X(t)} = \frac{\rho_\infty L}{2C}t^{-1/4},\label{eq:crit_rhom}\\
\delta\rho_+(t) & = \frac{C b_{1/4}\rho_\infty}{4\sqrt{\pi D}}t^{-1/4}.\label{eq:crit_drhop}
\end{align}

Using the force balance~(\ref{eq:force_bal_si}) we get
\begin{align}
f & = P(\rho(X_2^+))-P(\rho(X_2^-))\\
& = P(\rho_\infty + \delta\rho_+)-P(\rho_-)\\
& \simeq P(\rho_\infty) + \delta\rho_+P'(\rho_\infty) - \rho_- P'(0).
\end{align}
At the critical force, $f=P(\rho_\infty)$, hence
\begin{equation}
\delta\rho_+P'(\rho_\infty) = \rho_- P'(0).
\end{equation}

We can determine the value of $C$ with equations~(\ref{eq:crit_rhom}) and (\ref{eq:crit_drhop}):
\begin{equation}
C^2 = \frac{2\sqrt{\pi D}P'(0)L}{b_{1/4}P'(\rho_\infty)}.
\end{equation}
Finally,
\begin{align}
X(t)
& \sim \sqrt{\frac{2\sqrt{\pi}}{b_{1/4}}}\sqrt{\frac{P'(0)L}{P'(\rho_\infty)}} (Dt)^{1/4}\\
& \simeq 0.82\sqrt{\frac{P'(0)L}{P'(\rho_\infty)}} (Dt)^{1/4}.
\end{align}

\section{Two probes submitted to arbitrary forces}\label{ap:two_arb}

We consider two probes submitted to arbitrary forces $f_1$ and $f_2$.
We focus on the bound and unbound configurations.

\subsection{Bound configuration}\label{}

When the probes are bound, their velocities are $V_1(t)\sim V_2(t) \sim \tA\sqrt{D/t}$.
The density between the probes obeys an advection-diffusion equation in the reference frame of the probes, the advection being set by $V_i(t)$. 
Since the advection velocity decreases with time, diffusion dominates at long times and the density profile becomes uniform between the probes; we denote its value $\rho_1$.
The density at the right of probe 2 is $\rho_\infty g(\tA)$ and the density at the left of probe 1 is $\rho_\infty g(-\tA)$.

Here also, we can use the force balance (\ref{eq:force_bal_si}) :
\begin{align}
P(\rho_1) - P(\rho_\infty g(-\tA)) & = f_1,\\
P(\rho_\infty g(\tA))-P(\rho_1) & = f_2.
\end{align}
Summing these expressions, we get
\begin{equation}\label{eq:sum_forces_2p}
P(\rho_\infty g(\tA))-P(\rho_\infty g(-\tA)) = f_1+f_2 = F:
\end{equation}
this is the equation giving $\tA$ for a single probe submitted to a force $F$.
This result means that two bound probes behave like a single probe.

This solution is valid as long as $f_1>-P(\rho_\infty g(-\tA))$ and $f_2< P(\rho_\infty g(\tA))$.
From equation (\ref{eq:sum_forces_2p}), we see that these conditions are equivalent.

\subsection{Unbound configuration}\label{}

When the probes unbind, their velocties are of the form $V_i(t)\sim \tA_i\sqrt{D/t}$ with $\tA_1<\tA_2$. 
The density between them decays to zero, and the densities at the left of probe 1 and at the right of probe 2 are
\begin{align}
\rho(X_1^-) & = \rho_\infty g(-\tA_1),\\
\rho(X_2^+) & = \rho_\infty g(\tA_2).
\end{align}
As in the previous situations, the force balance (\ref{eq:force_bal_si}) still holds, leading to 
\begin{align}
-P(\rho_\infty g(-\tA_1)) & = f_1,\\
P(\rho_\infty g(\tA_2)) & = f_2.
\end{align}
These equations give $\tA_1$ and $\tA_2$.

\section{Theoretical predictions for continuous systems}\label{ap:th_cont}

We show briefly how our results can be extended to continuous systems with arbitrary short-range interactions with overdamped dynamics.
We are interested in two specific interactions. 
The first one is the hard-rod interaction (this is the Tonks gas), which corresponds to the experiments of Ref.~\cite{Lin2005}.
The pressure of the gas of hard rods with length $a$ is
\begin{equation}
P\ind{HR}(\rho) = \frac{\kT\rho}{1-a\rho}
\end{equation}

The second is the dipolar interaction which corresponds to the potential $U(r)=A/r^3$; it is induced between paramagnetic colloids with a magnetic field in Ref.~\cite{Wei2000}.
The pressure of this gas is not known exactly, but it can be approximated by a virial expansion at low density:
\begin{equation}
P\ind{dip}(\rho) \simeq kT\rho \left[1  + 1.35 a\rho + 1.40 (a\rho)^2\right],
\end{equation}
where $a=[A/(\kT)]^{1/3}$ is the caracteristic scale associated with the interaction.

The dynamics of these systems is not diffusive, but involves a density-dependent collective diffusion coefficient $D(\rho)$:
\begin{equation}
\frac{\partial \rho}{\partial t}(x,t)=\frac{\partial}{\partial x}\left[D(\rho(x,t))\frac{\partial \rho}{\partial x}(x,t) \right].
\end{equation}
In absence of hydrodynamic interactions, the collective diffusion coefficient is given by 
\begin{equation}
D(\rho)=\kappa_0 P'(\rho),
\end{equation}
where $\kappa_0$ is the individual mobility of the particles~\cite{Pusey1975, Lin2005}.

When the collective diffusion coefficient depends on the density, the result of Sec.~\ref{sub:sol_diff} does not apply when the probe moves rapidly, i.e., as $A\sqrt{t}$, with $A$ of order 1.
However, when the probe moves slowly, either as $A\sqrt{t}$ with $A\ll 1$ or as $t^{1/4}$, the density of the bath is only weakly perturbed, and the result of Secs.~\ref{sub:sol_diff}, \ref{sub:critical} can be applied with $D=D(\rho_\infty)$.

As a consequence, with two probes submitted to opposite forces, the bound regime and the critical 
regime are described by the equations given in the main text.
This is used to give the theoretical predictions shown in Fig.~\ref{fig:tonks_dipolar}.

 \end{document}